\documentclass[twocolumn]{aastex61}

\submitjournal{ApJL}

\shorttitle{Expansion of Solar Wind}
\shortauthors{Garton et al.}

\begin{document}

\title{Expansion of High Speed Solar Wind Streams From Coronal Holes Through the Inner Heliosphere}

\email{gartont@tcd.ie}

\author{Tadhg M. Garton}
\affil{School of Physics, \\
Trinity College Dublin, Dublin 2 \\
Ireland}
\affil{School of Cosmic Physics, \\
Dublin Institute of Advanced Studies, Dublin 2 \\
Ireland}

\author{Sophie A. Murray}
\affiliation{School of Physics, \\
Trinity College Dublin, Dublin 2 \\
Ireland}
\affil{School of Cosmic Physics, \\
Dublin Institute of Advanced Studies, Dublin 2 \\
Ireland}

\author{Peter T. Gallagher}
\affil{School of Cosmic Physics, \\
Dublin Institute of Advanced Studies, Dublin 2 \\
Ireland}
\affiliation{School of Physics, \\
Trinity College Dublin, Dublin 2 \\
Ireland}



\begin{abstract}

Coronal holes (CHs) are regions of open magnetic flux which are the source of high speed solar wind (HSSW) streams. To date, it is not clear which aspects of CHs are of most influence on the properties of the solar wind as it expands through the Heliosphere. Here, we study the relationship between CH properties extracted from AIA (Atmospheric Imaging Assembly) images using CHIMERA (Coronal Hole Identification via Multi-thermal Emission Recognition Algorithm) and HSSW measurements from ACE (Advanced Composition Explorer) at L1. For CH longitudinal widths $\Delta\theta_{CH}<$67$^{\circ}$, the peak SW velocity ($v_{max}$) is found to scale as $v_{max}~\approx~330.8~+~5.7~\Delta\theta_{CH}$~km~s$^{-1}$. For larger longitudinal widths ($\Delta\theta_{CH}>$67$^{\circ}$), $v_{max}$ is found to tend to a constant value ($\sim$710~km~s$^{-1}$). Furthermore, we find that the duration of HSSW streams ($\Delta t$) are directly related to the longitudinal width of CHs ($\Delta t_{SW}$~$\approx$~0.09$\Delta\theta_{CH}$) and that their longitudinal expansion factor is $f_{SW}~\approx 1.2~\pm 0.1$. We also derive an expression for the coronal hole flux-tube expansion factor, $f_{FT}$, which varies as $f_{SW} \gtrsim f_{FT} \gtrsim 0.8$. These results enable us to estimate the peak speeds and durations of HSSW streams at L1 using the properties of CHs identified in the solar corona. 

\end{abstract}




 \section{Introduction} \label{sec:intro}

Coronal holes (CHs) are low density regions of open magnetic field which appear dark in EUV wavelengths and are known to be associated with the acceleration of high-speed solar wind (HSSW) streams \citep{Krieger73, Cranmer02, Tu05, Cranmer09}. Due to their slow evolution, CHs can exist for a number of solar rotations, ranging in lifetime from months to years \citep{Timothy75, Bohlin77, deToma11, Krista11}. This slow evolution allows for relatively precise forecasting of HSSW streams emanating from CHs, and their potential trajectory through interplanetary space to 1~AU \citep{Heinemann18}. 

The solar wind is a stream of charged particles, largely protons and electrons, traveling outward from the Sun toward the edge of the heliosphere. This stream is classified into slow and fast variants, the former with typical speeds and temperatures of $\sim$400~km~s\textsuperscript{-1} and $\sim$10\textsuperscript{5}~K at 1~AU respectively \citep{Marsch06}, and the latter with speeds and temperatures of up to $\sim$780~km~s\textsuperscript{-1} and $\sim$10\textsuperscript{6}~K at 1~AU respectively \citep{Cranmer02}. The solar wind is of interest to operational space weather forecasters due to the potential damage it can cause to satellites through differential and bulk charging, and its association with geomagnetic storms and their impacts at Earth \citep{Boteler01, Huttunen08, Marshall12, Blake16}.

Empirical studies have shown that the properties of the HSSW, and hence their potential impacts, are largely governed by the properties of their originating CH regions \citep{Arge00,Vrsnak07,Rotter12}. Previous work has shown that HSSW velocity at 1AU is inversely proportional to the expansion of magnetic flux-tubes within the CH boundaries \citep{Levine77,Wang91} and can be estimated for a given CH through the Wang-Sheely (WS) model \citep{Wang90}. Magnetic flux-tube expansion can be described by a two-dimensional unitless comparison of magnetic flux density between two surfaces known as the magnetic flux-tube expansion factor \citep{Wang97}, as follows:

\begin{equation}
f(r,\theta)=\left(\frac{R_\odot}{r}\right)^{2}\frac{B_{r}(R_\odot,\theta_\odot,\phi_\odot)}{B_{r}(r,\theta,\phi)}
\label{eqn1}
\end{equation}

\noindent In this case, the expansion factor, $f$, is described between the solar surface and the source surface at radial distance $r$~=~2.5~$R_\odot$, where $B_{r}$ describes the magnetic field for a given surface, and $\theta$ and $\phi$ define longitude and latitude position information along the magnetic fields lines. \cite{Pinto17} simplified this expansion factor to a dimensionless comparison of the area, $A$, occupied by a flux-tube at two surface heights, $r_{\odot}$ and $r$, as follows:

\begin{equation}
f=\frac{A_{r}}{A_{\odot}}\left(\frac{r_{\odot}}{r}\right)^{2}
\label{eqn2}
\end{equation}

\noindent An alternate model for predicting HSSW speed proposed by \cite{Riley01} states that the SW velocity originating from a point within a CH boundary is positively correlated with the minimum Distance from the Coronal Hole Boundary (DCHB). This model has since been validated empirically from Ulysses measurements of polar CHs across 12 Carrington rotations by \cite{Riley03}. The WS and DCHB models have since been combined and improved to include real-time updating of an input magnetogram in the widely-used Wang-Sheeley-Arge (WSA) model \citep{Arge00, Arge04, Riley15}.

Previous observational studies have confirmed the inverse correlation between solar wind speed and magnetic flux-tube expansion. Notably, \cite{Wang90} performed a 22-year examination of the relationship between the solar wind and the rate of flux-tube expansion in the corona, which confirmed this inverse relationship and concluded that the WS model can be used to reproduce the overall patterns of fast and slow wind. This study was expanded by \cite{Wang97} using direct measurements from the Ulysses spacecraft. From these measurements the range of potential expansion factors and their associated wind speeds were estimated, however it was observed that the expansion factor model often over predicted very fast wind near the ecliptic plane. \cite{Pinto16} used a global magnetohydrodynamic simulation to confirm the speed of solar wind depends on the geometry of the open magnetic flux-tubes through which it flows. These findings were further used in \cite{Pinto17} to derive a three-dimensional model of the structure of the solar wind. In recent times many studies have investigated the link between the solar wind and EUV images of the solar corona. \cite{Temmer07} investigated periodicity in the presence of CH areas at central meridian and rises in SW speed. \cite{Vrsnak07} analyzed the relationship between coronal hole area/positions and physical characteristics of the associated high speed stream.

Here, an analysis of the longitudinal solar wind expansion is performed for high speed solar wind streams traveling through interplanetary space. Measurements of originating CH properties are estimated with images from the Solar Dynamics Observatory Atmospheric Imaging Assembly \citep[AIA;][]{Lemen12} and Helioseismic and Magnetic Imager \citep[HMI;][]{Scherrer12} using the Coronal Hole Identification via Multi-thermal Emission Recognition Algorithm (CHIMERA; \citeauthor{Garton17}~\citeyear{Garton17}). For the first time these CHIMERA CH properties are compared with \textit{in-situ} measurements of HSSW streams from the Advanced Composition Explorer (ACE; \citeauthor{Stone98}~\citeyear{Stone98}) at the L1 point. Specifically, comparisons are made between CH longitudinal width, observed when CHs are located at central meridian, and the duration of the HSSW stream produced as measured at L1, for the period 2016 to 2017. In this case of flux-tube expansion, the CH surface is defined to be inclusive of the magnetic fields caused by the interaction of the CH boundary to it's surrounding plasma and the HSSW surface is defined to be inclusive of the tail end of the HSSW stream typically composed of material originating from the interaction of CH boundaries. From this correlation analysis it is possible to estimate the longitudinal component of the HSSW stream expansion through interplanetary space, as described in Section~\ref{sec:meth}. 

\section{Observation and Analysis} \label{sec:meth}

\begin{figure}[ht!]
\epsscale{1.1}
\plotone{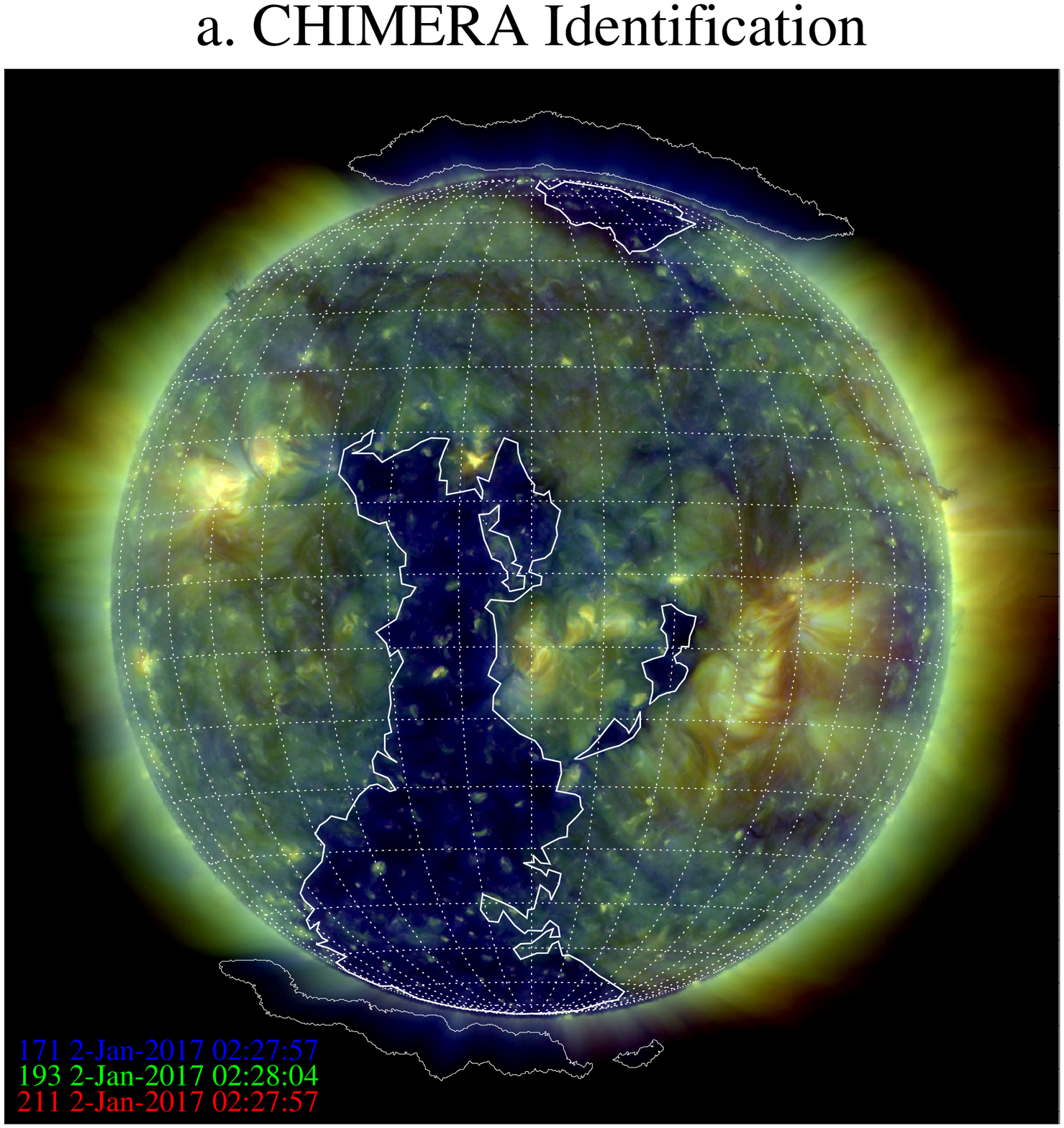}
\epsscale{1.1}
\plotone{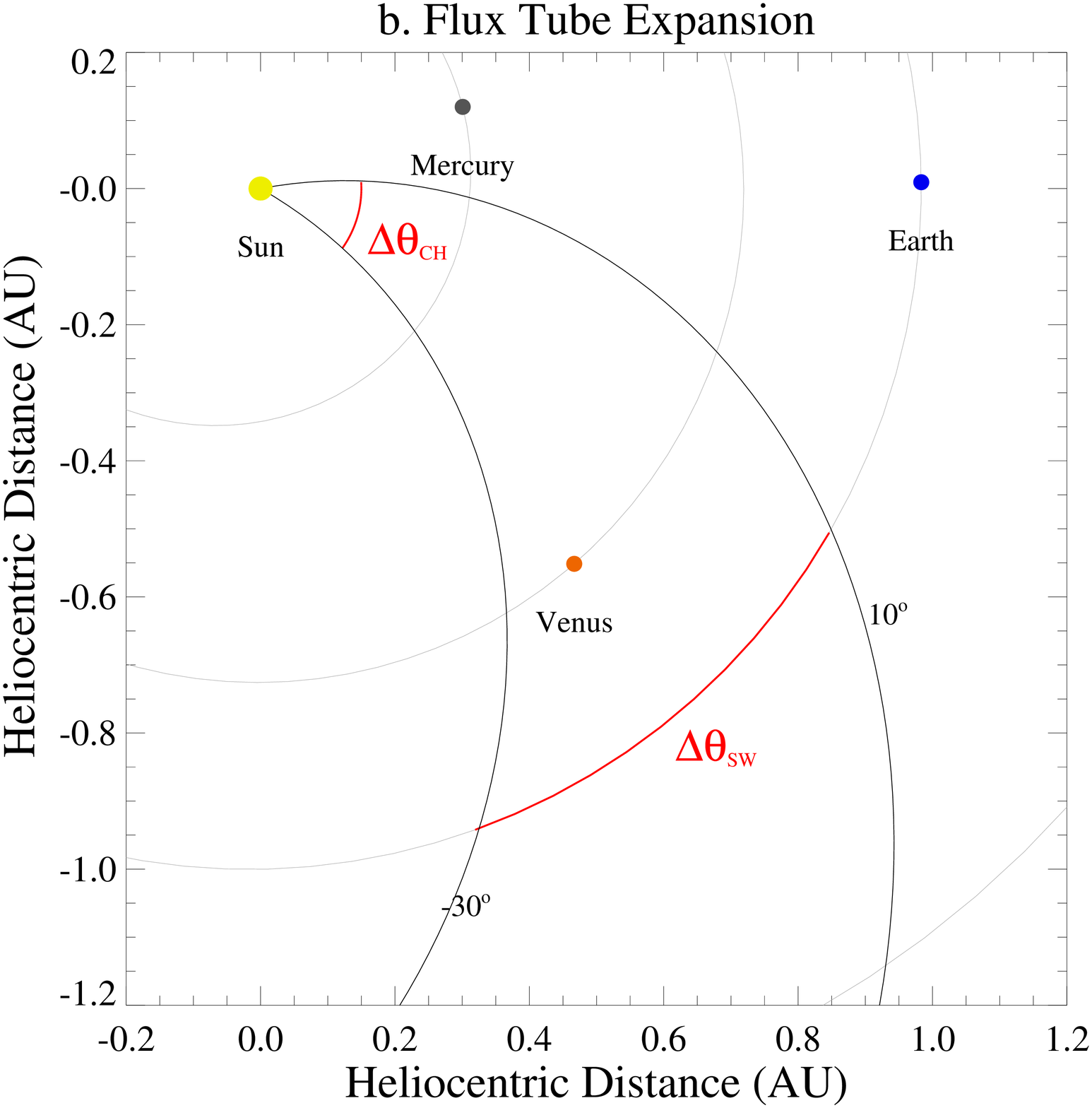}
\caption{(a.) Five on-disk CH regions on 2017 January 2 segmented by the CHIMERA algorithm on a tri-colour image of AIA 171\AA, 193\AA, and 211\AA. The largest CH on disk is located near the central meridian. Off-limb detected regions are classified as solar plumes caused by the extended open magnetic fields above CH regions, notably above the north and south polar CH regions. (b.) A ballistic model for HSSW stream propagation from the central on-disk CH. This simulation assumes the HSSW stream is emitted at $\sim$600~km~s$^{-1}$ from within the equatorial latitudes of the CH boundaries, -30$^\odot$ to +10$^\odot$. $\Delta\theta_{CH}$ and $\Delta\theta_{SW}$ denote the angular width of the CH on the solar disk and the HSSW stream at 1~AU respectively.} 
\label{fig:ch}
\end{figure}

\begin{figure}[!t]
\epsscale{1.2}
\plotone{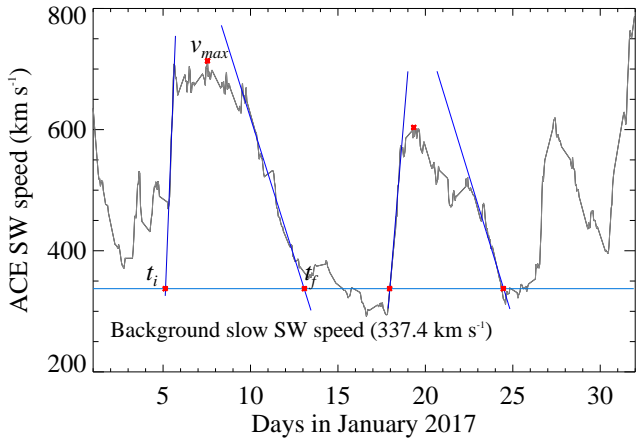}
\caption{Example start ($t_{i}$) and end ($t_{f}$) times for high-speed solar wind streams from ACE measurements of the solar wind speed for January 2017. These times are calculated as the point of intersection between a line fit to the inclining and declining phase of the solar wind stream with the mean background slow SW speed for that month. Maximum velocities ($v_{max}$) between these two times is considered as the peak wind velocity for that high speed stream.}
\label{meth}
\end{figure}

CHIMERA is an automatic CH identification and segmentation algorithm which extracts multiple property measurements from classified CH regions (see Figure~\ref{fig:ch}a), such as CH area, magnetic polarity, etc. The algorithm classifies these regions through a multi-thermal segmentation method, only accepting candidates that exhibit thermal and magnetic properties similar to that expected of a CH. Further details on the segmentation method can be found in \cite{Garton17}. Here, measurements of CH longitudinal width in degrees are extracted for CHs when their centroid is located closest to the central meridian. Analyzing values of CHs when centered at central meridian ensures a minimal loss of estimated width of extended CHs caused by occultation effects.

Since HSSW streams originating from CHs can vary in velocity from $\sim$400-800~km~s$^{-1}$ \citep{Cranmer02}, it can be assumed that the velocity of solar wind emitted within a CH boundary varies within this range \citep{Riley01,Riley03}. This variation of emitted speed implies that the angular width of HSSW streams can be extended should a particularly fast stream of solar wind be followed by a relatively slow stream still located within the CH boundary. The solar wind emitted from the Eastern and Western boundaries of the central CH from Figure~\ref{fig:ch}a is simulated using the HELiophysics Integrated Observatory (HELIO; \citeauthor{PSuarez12}~\citeyear{PSuarez12}) ballistic model in Figure~\ref{fig:ch}b.  The longitudinal component of solar wind expansion factor is derived in \cite{Krista12}, and can be defined from this figure as:

\begin{equation}
f_{SW}=\frac{\Delta\theta_{SW}}{\Delta\theta_{CH}}
\label{eqn3}
\end{equation}

The longitudinal width of the detected CH at 1R$_{\odot}$ is denoted by $\Delta\theta_{CH}$, while $\Delta\theta_{SW}$ indicates the longitudinal width of the HSSW at L1. For an expanding SW stream, $f>1.0$. Measurements of $\Delta\theta_{CH}$ are available through CHIMERA. Direct measurements of $\Delta\theta_{SW}$ are not currently possible, however it can be calculated from \textit{in-situ} measurements of HSSW stream duration taken from ACE, $\Delta t_{SW}$, using the angular velocity, $\Delta\theta_{SW} = \omega_{\odot}\Delta t_{SW}$. The solar wind's apparent angular velocity, $\omega_{\odot}$, is assumed to be equal to the synodic Carrington rotational velocity of the Sun. A comprehensive study by \cite{Oghrapishvili18} of CHs finds a variation of rotational velocities with latitude, with a plateau existing between $\pm$40$^{\circ}$. Here, only CHs associated with a measurable HSSW stream at L1 are analyzed, typically with some component of CH boundary existing within $\pm$40$^{\circ}$. Hence we assume a constant value of rotational velocity, $\sim$13.199$^{\circ}$day$^{-1}$. Hence, $f_{SW}$ was calculated using:

\begin{equation}
f_{SW}=\omega_{\odot}\frac{\Delta t_{SW}}{\Delta\theta_{CH}}
\label{eqn4}
\end{equation}

\noindent Equation~\ref{eqn4} can be rearranged to form a relation between the measurable parameters $\Delta t_{SW}$ and $\Delta\theta_{CH}$ as follows:

\begin{equation}
\Delta t_{SW}=\frac{f_{SW}}{\omega_{\odot}}\Delta\theta_{CH}
\label{eqn4.1}
\end{equation}

\noindent This relationship predicts the duration of upcoming HSSW streams from measurements of CH width at the central meridian, at 1R$_{\odot}$.

An example of ACE solar wind speed measurements for the month of January 2017 is shown in Figure~\ref{meth}. A line is fitted to the rising and declining phase of the stream in order to calculate the duration of a HSSW stream originating from a single CH. The intersections of these linear fits with the mean background slow solar wind speed for a given month, 337.4~km~s$^{-1}$ in January 2017, defines the start time ($t_{i}$) of the HSSW streams arrival at L1 and end time ($t_{f}$) of the HSSW streams interaction at L1. The difference between the start and end times of the stream is then calculated as, $\Delta~t_{SW}~=~t_{f}~-~t_{i}$. This method of calculating $\Delta t_{SW}$ removes potential errors caused by overlapping HSSW streams originating from closely-clustered CHs. This method estimates the behavior of the solar wind in the absence of perturbations caused by other solar features. An example of this occurrence is visible in Figure~\ref{meth} between 2017 January 2-5, where two peaks in solar wind velocity exist. These peaks are due to two detected CH regions located near to the large CH at central meridian, as shown in Figure~\ref{fig:ch}a.

By comparing the measurements of HSSW duration and width of their respective CHs, it is possible to draw a correlation and extract an estimation of the longitudinal expansion that HSSW streams undergo between their origin on the solar surface and their detection at L1.

\begin{figure}[!t]
\epsscale{1.2}
\plotone{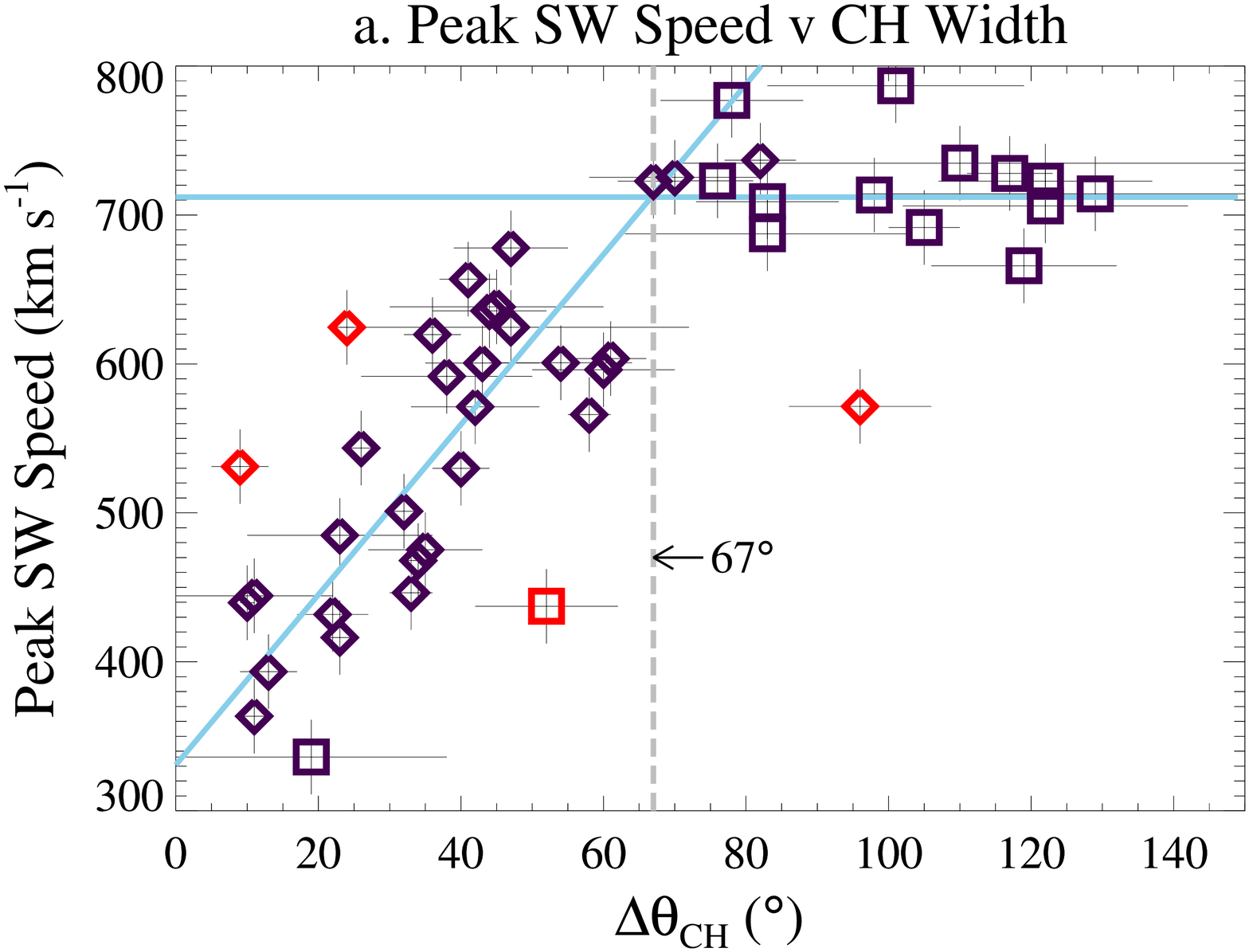}
\epsscale{1.2}
\plotone{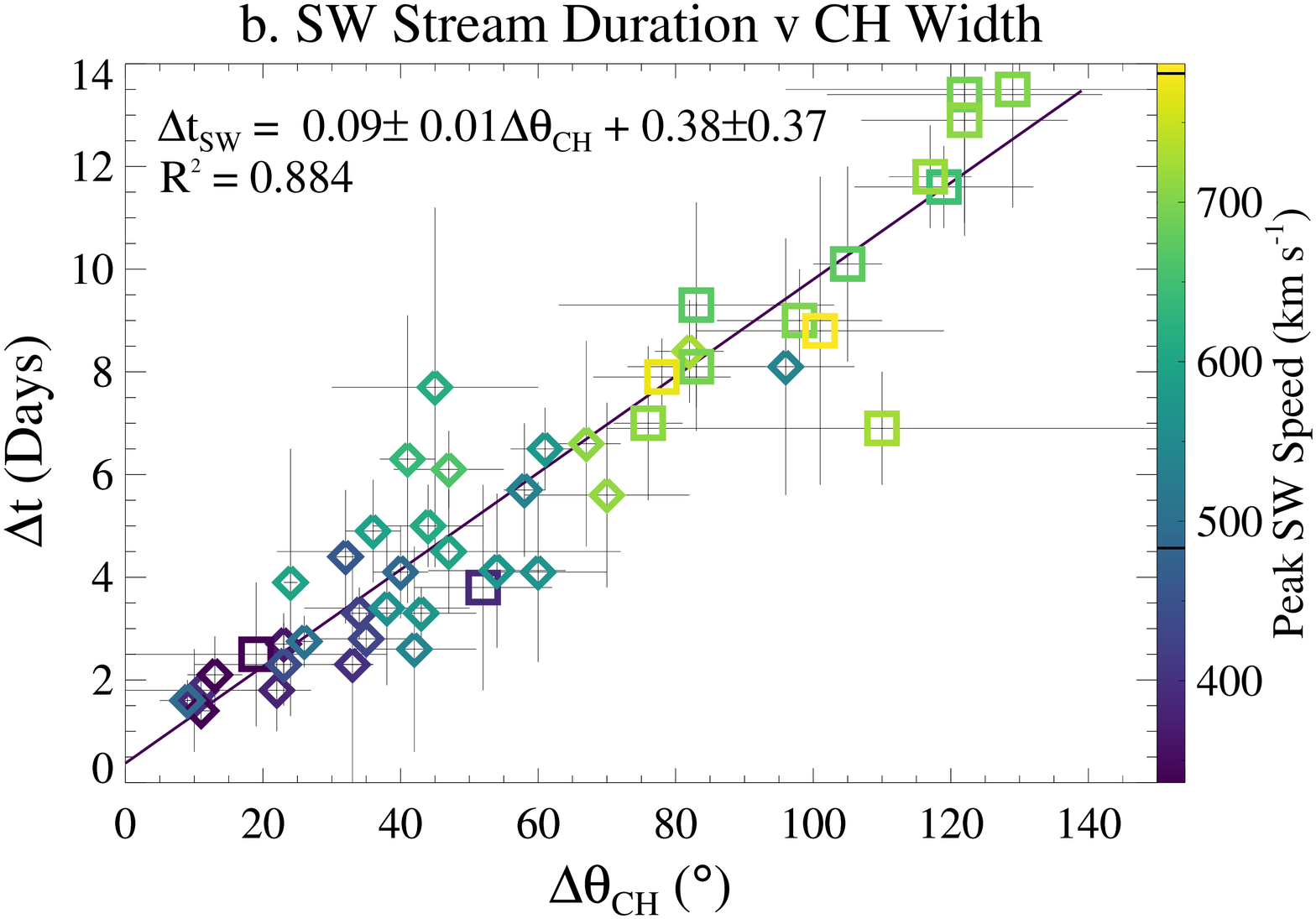}
\caption{(a.) Peak solar wind speed versus longitudinal width of detected CHs. A linear fit to both the high width and low width CH clusters is illustrated in blue. Significant outliers are highlighted in red and are due to irregularly shaped CHs with width and heights being significantly different. (b.) Correlation between the duration of HSSW streams to CH width. Symbol colour represents peak SW speed observed during the HSSW stream and scales from purple, $\sim$400~km~s$^{-1}$, to yellow, $\sim$700~km~s$^{-1}$. From this correlation, and Equation~\ref{eqn4}, it is possible to estimate the longitudinal expansion of HSSW streams through interplanetary space to be f$_{LON}\approx$~1.2$\pm$0.1. Squares represent CHs that connect to either of the solar magnetic poles and diamonds represent CHs with no connection to polar regions.}
\label{comp}
\end{figure}

\section{Results} \label{sec:res}

A comparison of CH width and associated SW peak velocity for CHs detected by CHIMERA at disk center during 2016 and 2017 is displayed in Figure~\ref{comp}a. Here, significant outliers are highlighted in red. These outliers are caused by irregularly shaped, extended CHs in the instance of non-polar CHs, diamonds, and by a possible near miss caused by a high latitude CH in the polar CH instance, square. Two linear relations between CH width and peak SW velocity are drawn. For CH regions of width $\lesssim$70$^{\circ}$ a relation of $v_{max}~\approx~330.8(\pm16.6)~+~5.7(\pm0.5)~\Delta\theta_{CH}$ is fit. This relation tapers off above $\approx70^{\circ}$ to a near constant speed of $\sim$710~km~s$^{-1}$, with a standard deviation of $\sim$50~km~s$^{-1}$. The intersection of these two regimes occurs at $\sim$67$^{\circ}~\pm$~11$^{\circ}$.

CH width and HSSW stream duration are compared in Figure~\ref{comp}b. Here, colour represents the peak SW speeds for the CH and ranges from purple, $\sim$400~km~s$^{-1}$, to yellow, $\sim$700~km~s$^{-1}$. Symbol shape describes CH topology, with squares representing CHs that link to either magnetic pole and diamonds represent non-polar related CH regions. The strong correlation between width and duration is demonstrated by the best fit line $\Delta t_{SW}$(Days)~$\approx~0.09~\pm~0.01~\Delta\theta_{CH}$(Deg), with a high $R^{2}$ value of 0.884. This high $R^{2}$ value may be due to 2016 having a larger number of very extended CHs than is typical. This relation enables the prediction of the durations of HSSW streams at Earth. Figure~\ref{comp} was replicated for area-based estimations from which a similar trend to Figure~\ref{comp}a was found for the observed dates, with a slightly better fit, however more outliers were apparent in the equivalent Figure~\ref{comp}b when comparing to duration. From the slope of this best fit linear relation and Equation~\ref{eqn4.1} it is possible to estimate the average longitudinal solar wind expansion factor using:

\begin{equation}
\frac{f_{SW}}{\omega_{\odot}}=0.09\pm0.01
\label{eqn5}
\end{equation}

\noindent Assuming this angular velocity is equal to that of the synodic Carrington rotation, 13.199$^{\circ}$day$^{-1}$, a general longitudinal expansion factor of $f_{SW}=1.2\pm0.1$ is obtained. This value implies HSSW streams will expand longitudinally while traveling through interplanetary space.

\begin{figure}[!t]
\epsscale{0.9}
\plotone{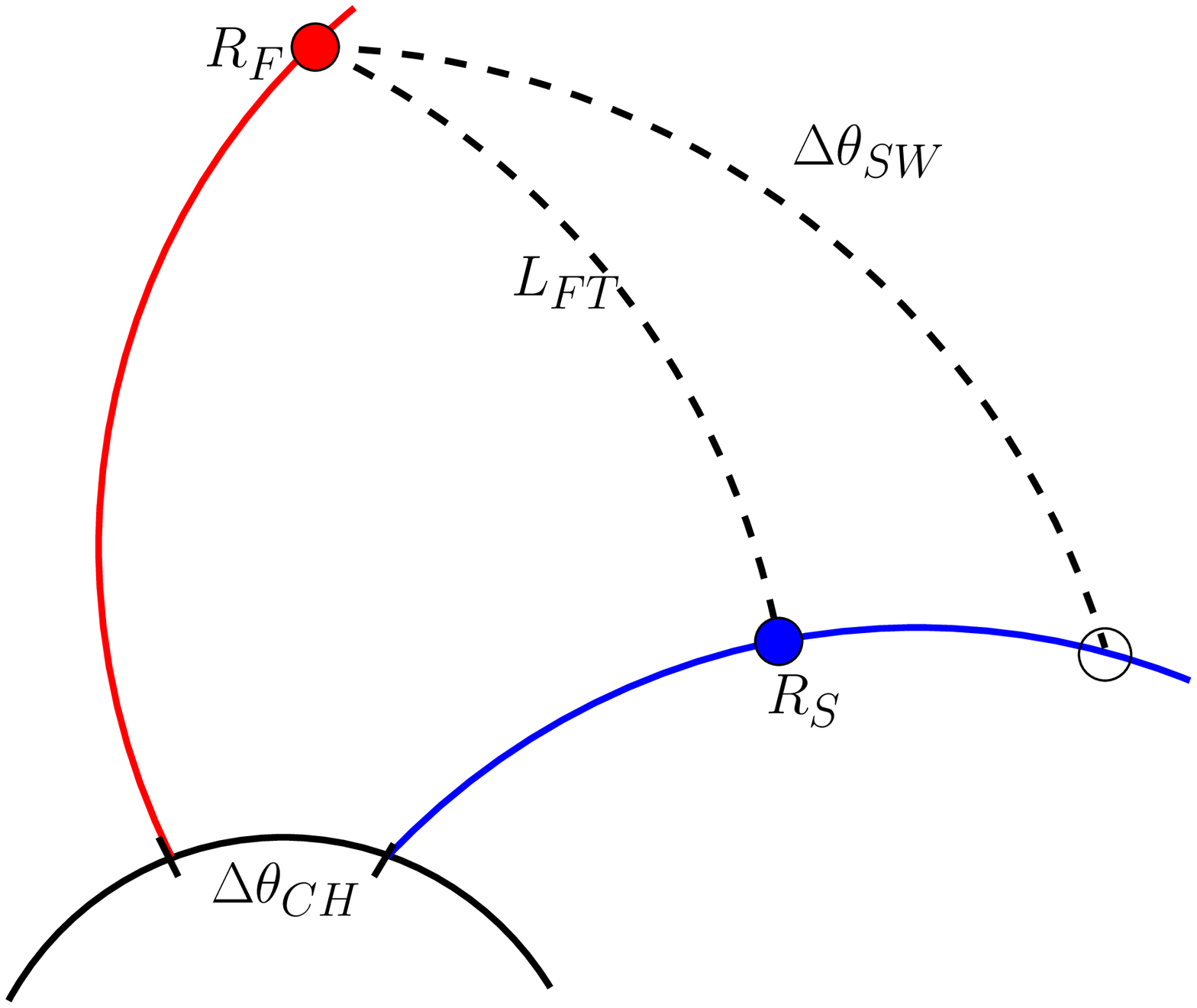}
\epsscale{1.}
\plotone{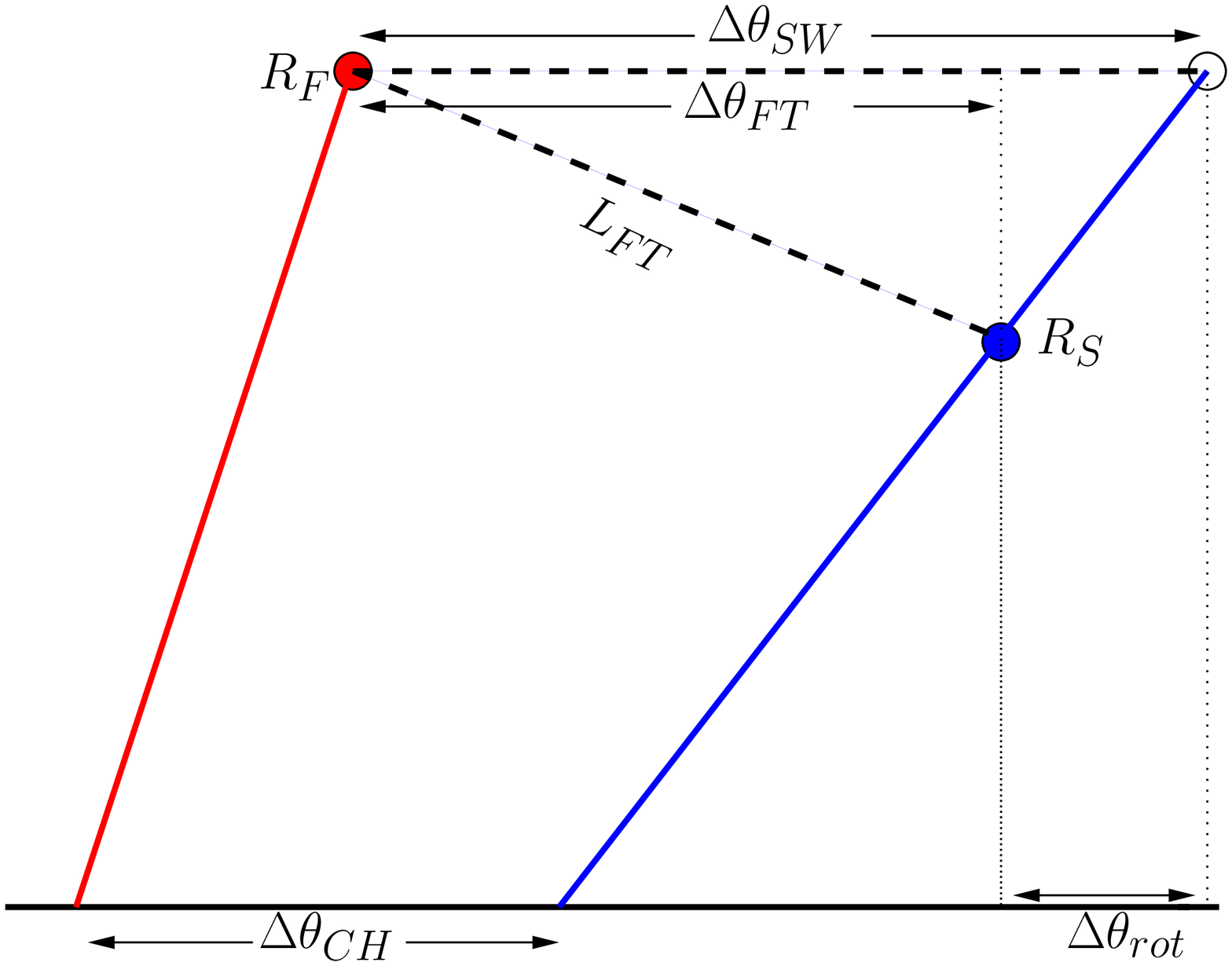}
\caption{(a.) Projection of a HSSW stream with a leading boundary (red) originating from the western CH boundary and a trailing boundary (blue) originating from the eastern CH boundary. $R_{F}$ and $R_{S}$ show the distance traveled by plasma packets emitted at the same time but traveling at different velocities. $\Delta\theta_{SW}$ represent the path of Earth through the HSSW stream and $L_{FT}$ approximates the continuous line of plasma packages, emitted from the Sun at the same time, when the leading boundary reaches some $R_{F}$ (b.) An identical projection of the HSSW in polar space. $\Delta\theta_{rot}$ here represents the angular difference between the trailing boundary of the HSSW at $R_{F}$ and $R_{S}$ and $\Delta\theta_{FT}$ is the longitudinal angular width of the HSSW flux-tube projected out to $R_{F}$.}
\label{fderiv}
\end{figure}
Due to the varying velocity profile across a CH, this calculation of $f_{SW}$ includes an additional component caused by the extra time for the relatively slower solar wind emitted from the eastern boundary of the CH to reach L1. This effect can be seen in Figure~\ref{fderiv} where the leading boundary of the HSSW reaches a distance of $R_{F}$, while the trailing boundary only reaches a boundary of $R_{S}$ such that $R_{S}=R_{F}(v_{S}/v_{F})$, where $v_{F}$ and $v_{S}$ are the velocities of the solar wind at the leading and trailing boundaries respectively. By correcting for this velocity variation across the HSSW stream, it is possible to estimate the longitudinal expansion of the CH flux-tube from the corona to L1 as follows:

\begin{equation}
f_{FT}^{long}=\frac{\Delta\theta_{FT}}{\Delta\theta_{CH}}=\frac{\Delta\theta_{SW}-\Delta\theta_{rot}}{\Delta\theta_{CH}}
\label{f-deriv-1}
\end{equation}

\noindent where $\Delta\theta_{rot}$ can be calculated from Figure~\ref{fderiv} as the angle the Sun has rotated in the time taken for the slow boundary to reach a distance $R_{F}$ traveling radially from $R_{S}$, $\Delta\theta_{rot}=\omega_{\odot}R_{F}[(1/v_{S})-(1/v_{F})]$ and $\Delta\theta_{SW}$ can be expressed as $f_{SW}\Delta\theta_{CH}$. Hence, $f_{FT}^{long}$ can be written as:

\begin{equation}
f_{FT}^{long}=f_{SW}-\frac{\omega_{\odot}R_{F}}{\Delta\theta_{CH}}\left(\frac{1}{v_{S}}-\frac{1}{v_{F}}\right)
\label{f-deriv-2}
\end{equation}

Furthermore, it is possible to estimate the width expansion of the overall open magnetic field of the coronal hole, which will henceforth be referred to as the coronal hole flux-tube expansion factor ($f_{FT}$):

\begin{equation}
f_{FT}=\frac{L_{FT}}{r}\left(\frac{r_{CH}}{l_{CH}}\right)=\frac{1}{\Delta\theta_{CH}}\int_{0}^{L_{FT}}\frac{dL_{FT}}{r}
\label{f-deriv-3}
\end{equation}

\noindent where the length $L_{FT}$ is approximated as a segment of a spiral such that:

\begin{equation}
f_{FT}=\frac{1}{\Delta\theta_{CH}}\int_{0}^{\Delta\theta_{FT}}\frac{\sqrt{r^{2}+(\partial r/\partial\theta)^2}}{r}d\theta
\label{f-deriv-4}
\end{equation}

\noindent The radius of a given spiral changes as a function of $\theta$, in this case calculated from Figure~\ref{fderiv} as:

\begin{equation}
r=R_{S}+\frac{(R_{F}-R_{S})}{\Delta\theta_{FT}}\theta
\label{f-deriv-5}
\end{equation}

\noindent Hence, Equation~\ref{f-deriv-4} can be simplified to:

\begin{equation}
f_{FT}=\frac{1}{\Delta\theta_{CH}}\int_{0}^{\Delta\theta_{FT}}\sqrt{1+\left(\frac{\Delta v}{v_{S}\Delta\theta_{FT}+\theta\Delta v}\right)^{2}}d\theta
\label{f-deriv-6}
\end{equation}

\noindent where $\Delta v$ is the difference of velocities between the leading and trailing HSSW stream boundary, ($v_{F}-v_{S}$). Integrating gives the general equation for the coronal hole flux-tube expansion factor as:

\begin{equation}
f_{FT}=\frac{\alpha-\beta}{\Delta v\Delta\theta_{CH}}+\frac{1}{2\Delta\theta_{CH}}ln\left(\frac{[\beta+\Delta v][\alpha-\Delta v]}{[\beta-\Delta v][\alpha+\Delta v]}\right)
\label{f-deriv-7}
\end{equation}

\noindent where

\begin{equation}
\alpha=\sqrt{(v_{F}\Delta\theta_{FT})^{2}+\Delta v^{2}}
\label{f-deriv-8}
\end{equation}

\begin{equation}
\beta=\sqrt{(v_{S}\Delta\theta_{FT})^{2}+\Delta v^{2}}
\label{f-deriv-9}
\end{equation}

\noindent where $\Delta\theta_{FT}$ is the angular width of the flux-tube ($\Delta\theta_{FT}=\Delta\theta_{SW}-\Delta\theta_{rot}$). From these derivations it is possible to estimate the range of possible expansion factors. From empirical measurements $f_{SW}$ will remain at 1.2 regardless of CH width. $f_{FT}^{long}$ will range from 1.2 for small CHs to ${\sim}0.5$ for $\Delta\theta_{CH}\approx60^{\circ}$. Above $\Delta\theta_{CH}\approx60^{\circ}$, $f_{FT}^{long}$ tends towards a constant value of 1.

$f_{FT}$ is undefined for ${\Delta}v=0$, however, due to the correlation between ${\Delta}v$ and $\Delta\theta_{CH}$, ${\Delta}v$ is zero only when $\Delta\theta_{CH}=0$, i.e., when no CH is present. Hence, Equations~\ref{f-deriv-7}-\ref{f-deriv-9} only apply when both ${\Delta}v>0$ and $\Delta\theta_{CH}>0$, i.e., when a CH is present.  $f_{FT}$ approaches a value of 1.2 for very small CHs, $\lim_{\Delta\theta_{CH}{\to}0} f_{FT}(\Delta\theta_{CH})=1.2$, and as $\Delta\theta_{CH}$ increases to a small width CH (${\sim}20^{\circ}$), $f_{FT}$ approaches a value of ${\sim}0.8$. Above this CH width, values of $f_{FT}$ tend towards ${\sim}1$.

\section{Discussion and Conclusions} \label{sec:conc}

Here, the relationship between CH width, a CH property made available by the CHIMERA algorithm, and the properties of the associated solar wind measured at L1 by the ACE satellite has been investigated. The results show that a positive correlation exists between the peak SW speed of HSSW stream and the width of their originating CHs for widths $\lesssim$67$^{\circ}$. Variations from a direct correlation are due to the HSSW speed being related to the area of CH regions, which varies independently from longitudinal width, and possible near misses of HSSW streams. Furthermore, other CH properties likely have a further contribution to the solar wind speed. Above $\sim67^{\circ}$ width the peak SW velocity appears to become constant at $\sim$710~km~s$^{-1}$ regardless of CH width, with a standard deviation of $\sim$50~km~s$^{-1}$. These speeds are consistent with the theory of HSSW streams emanating from CH regions by \cite{Cranmer09}. Furthermore, this relation is similar to the relation between HSSW velocity and distance from a coronal boundary found by \cite{Riley03}.

From the strong correlation of HSSW stream duration to CH width in Figure~\ref{comp}b it is clear these properties are fundamentally linked. Hence it is possible to predict the duration of an incoming stream of HSSW using the best fit linear relation, $\Delta t_{SW}~=~0.09(\pm0.01)~\Delta\theta_{CH}~+~0.38(\pm0.37)$. Combined with empirical measurements, such as in \cite{Vrsnak07}, and the expanded study by \cite{Verbanac11}, a prediction of the start and end time of a HSSW streams interaction with Earth is possible.
 
From these measurements of CH width and stream duration we calculate an average longitudinal solar wind expansion factor of 1.2~$\pm$~0.1. This value implies the HSSW always expands longitudinally from 1$R_{\odot}$ to 1AU. This consistent expansion is likely a composite of the HSSW flux-tube expanding and an increased longitudinal width caused by differing arrival times of the leading and trailing boundaries at 1AU. By correcting for this variation in arrival times, it is possible to estimate the projected longitudinal expansion of the HSSW flux-tube at $R_{F}$ from Equation~\ref{f-deriv-2}, which ranges from $f_{SW}~\gtrsim~f_{FT}^{long}~\gtrsim~0.5$. Then, by approximating the structure of the flux-tube as a spiral, it is possible to estimate the coronal hole flux-tube expansion factor from Equations~\ref{f-deriv-6}~and~\ref{f-deriv-7}, which ranges from $f_{SW}~\gtrsim~f_{FT}~\gtrsim~0.8$. These values of flux-tube expansion are very low compared to empirical area flux-tube expansion values found by \cite{Wang97} of ${<}3.5$ to ${>}18$, or modeled values by \cite{Pinto17} from 1 to $\sim$100. This discrepancy is likely due to the focus here on the longitudinal flux-tube expansions and the potential of flux-tubes expanding non-uniformly in the longitudinal and latitudinal directions. Furthermore, previous studies have focused on the expansion of flux-tubes originating in polar CH regions or the expansion of individual magnetic funnels within a CH boundary, as in the \citeauthor{Pinto17} work. This work instead averages the expansion factors of all magnetic funnels within the CHs anywhere on the solar disk that correlates with geomagnetic storm activity.

These average values of $f_{SW}$ determined here are useful for operational space weather forecasting efforts, for the first time enabling a prediction of the duration and max speeds of HSSW streams and the expansion of the HSSW flux-tubes merely from an estimation of longitudinal width of CH regions. These results demonstrate an example of the potential connections that can be discovered between CHs and the solar wind using the new automated CHIMERA method.

\begin{acknowledgements}
T.~M.~G. is supported by a Government of Ireland Studentship from the Irish Research Council (IRC). S.~A.~M. is supported by the Irish Research Council Postdoctoral Fellowship Programme and the Air Force Office of Scientific Research award number FA9550-17-1-039. Images used for this research are constructed from images courtesy of NASA/SDO and the AIA, EVE, and HMI science teams. We thank the ACE SWEPAM instrument team and the ACE Science Center for providing the ACE data. We thank the anonymous referee for their constructive suggestions to improve the manuscript.
\end{acknowledgements}


\begin{thebibliography}{}
\expandafter\ifx\csname natexlab\endcsname\relax\def\natexlab#1{#1}\fi
\providecommand{\url}[1]{\href{#1}{#1}}

\bibitem[{{Arge} {et~al.}(2004){Arge}, {Luhmann}, {Odstrcil}, {Schrijver}, \&
  {Li}}]{Arge04}
{Arge}, C.~N., {Luhmann}, J.~G., {Odstrcil}, D., {Schrijver}, C.~J., \& {Li},
  Y. 2004, Journal of Atmospheric and Solar-Terrestrial Physics, 66, 1295

\bibitem[{{Arge} \& {Pizzo}(2000)}]{Arge00}
{Arge}, C.~N., \& {Pizzo}, V.~J. 2000, \jgr, 105, 10465

\bibitem[{{Blake} {et~al.}(2016){Blake}, {Gallagher}, {McCauley}, {Jones},
  {Hogg}, {Campany{\`a}}, {Beggan}, {Thomson}, {Kelly}, \& {Bell}}]{Blake16}
{Blake}, S.~P., {Gallagher}, P.~T., {McCauley}, J., {et~al.} 2016, Space
  Weather, 14, 1136

\bibitem[{{Bohlin}(1977)}]{Bohlin77}
{Bohlin}, J.~D. 1977, \solphys, 51, 377

\bibitem[{{Boteler}(2001)}]{Boteler01}
{Boteler}, D.~H. 2001, Washington DC American Geophysical Union Geophysical
  Monograph Series, 125, 347

\bibitem[{{Cranmer}(2002)}]{Cranmer02}
{Cranmer}, S.~R. 2002, \ssr, 101, 229

\bibitem[{Cranmer(2009)}]{Cranmer09}
Cranmer, S.~R. 2009, Living Reviews in Solar Physics, 6, 3

\bibitem[{{de Toma}(2011)}]{deToma11}
{de Toma}, G. 2011, \solphys, 274, 195

\bibitem[{{Garton} {et~al.}(2017){Garton}, {Gallagher}, \& {Murray}}]{Garton17}
{Garton}, T.~M., {Gallagher}, P.~T., \& {Murray}, S.~A. 2017, ArXiv e-prints,
  arXiv:1711.11476

\bibitem[{{Heinemann} {et~al.}(2018){Heinemann}, {Temmer}, {Hofmeister},
  {Veronig}, \& {Vennerstr{\o}m}}]{Heinemann18}
{Heinemann}, S.~G., {Temmer}, M., {Hofmeister}, S.~J., {Veronig}, A.~M., \&
  {Vennerstr{\o}m}, S. 2018, \apj, 861, 151

\bibitem[{{Huttunen} {et~al.}(2008){Huttunen}, {Kilpua}, {Pulkkinen},
  {Viljanen}, \& {Tanskanen}}]{Huttunen08}
{Huttunen}, K.~E.~J., {Kilpua}, S.~P., {Pulkkinen}, A., {Viljanen}, A., \&
  {Tanskanen}, E. 2008, Space Weather, 6, S10002

\bibitem[{{Krieger} {et~al.}(1973){Krieger}, {Timothy}, \&
  {Roelof}}]{Krieger73}
{Krieger}, A.~S., {Timothy}, A.~F., \& {Roelof}, E.~C. 1973, \solphys, 29, 505

\bibitem[{{Krista}(2012)}]{Krista12}
{Krista}, L.~D. 2012, PhD thesis, PhD Thesis, 2012

\bibitem[{{Krista} {et~al.}(2011){Krista}, {Gallagher}, \&
  {Bloomfield}}]{Krista11}
{Krista}, L.~D., {Gallagher}, P.~T., \& {Bloomfield}, D.~S. 2011, \apjl, 731,
  L26

\bibitem[{{Lemen} {et~al.}(2012){Lemen}, {Title}, {Akin}, {Boerner}, {Chou},
  {Drake}, {Duncan}, {Edwards}, {Friedlaender}, {Heyman}, {Hurlburt}, {Katz},
  {Kushner}, {Levay}, {Lindgren}, {Mathur}, {McFeaters}, {Mitchell}, {Rehse},
  {Schrijver}, {Springer}, {Stern}, {Tarbell}, {Wuelser}, {Wolfson}, {Yanari},
  {Bookbinder}, {Cheimets}, {Caldwell}, {Deluca}, {Gates}, {Golub}, {Park},
  {Podgorski}, {Bush}, {Scherrer}, {Gummin}, {Smith}, {Auker}, {Jerram},
  {Pool}, {Soufli}, {Windt}, {Beardsley}, {Clapp}, {Lang}, \&
  {Waltham}}]{Lemen12}
{Lemen}, J.~R., {Title}, A.~M., {Akin}, D.~J., {et~al.} 2012, \solphys, 275, 17

\bibitem[{{Levine} {et~al.}(1977){Levine}, {Altschuler}, \&
  {Harvey}}]{Levine77}
{Levine}, R.~H., {Altschuler}, M.~D., \& {Harvey}, J.~W. 1977, \jgr, 82, 1061

\bibitem[{{Marsch}(2006)}]{Marsch06}
{Marsch}, E. 2006, Living Reviews in Solar Physics, 3, 1

\bibitem[{{Marshall} {et~al.}(2012){Marshall}, {Dalzell}, {Waters},
  {Goldthorpe}, \& {Smith}}]{Marshall12}
{Marshall}, R.~A., {Dalzell}, M., {Waters}, C.~L., {Goldthorpe}, P., \&
  {Smith}, E.~A. 2012, Space Weather, 10, S08003

\bibitem[{{Oghrapishvili} {et~al.}(2018){Oghrapishvili}, {Bagashvili},
  {Maghradze}, {Gachechiladze}, {Japaridze}, {Shergelashvili},
  {Mdzinarishvili}, \& {Chargeishvili}}]{Oghrapishvili18}
{Oghrapishvili}, N.~B., {Bagashvili}, S.~R., {Maghradze}, D.~A., {et~al.} 2018,
  Advances in Space Research, 61, 3039

\bibitem[{{P{\'e}rez-Su{\'a}rez} {et~al.}(2012){P{\'e}rez-Su{\'a}rez},
  {Maloney}, {Higgins}, {Bloomfield}, {Gallagher}, {Pierantoni}, {Bonnin},
  {Cecconi}, {Alberti}, {Bocchialini}, {Dierckxsens}, {Opitz}, {Le Blanc},
  {Aboudarham}, {Bentley}, {Brooke}, {Coghlan}, {Csillaghy}, {Jacquey},
  {Lavraud}, \& {Messerotti}}]{PSuarez12}
{P{\'e}rez-Su{\'a}rez}, D., {Maloney}, S.~A., {Higgins}, P.~A., {et~al.} 2012,
  \solphys, 280, 603

\bibitem[{{Pinto} {et~al.}(2016){Pinto}, {Brun}, \& {Rouillard}}]{Pinto16}
{Pinto}, R.~F., {Brun}, A.~S., \& {Rouillard}, A.~P. 2016, \aap, 592, A65

\bibitem[{{Pinto} \& {Rouillard}(2017)}]{Pinto17}
{Pinto}, R.~F., \& {Rouillard}, A.~P. 2017, \apj, 838, 89

\bibitem[{{Riley} {et~al.}(2015){Riley}, {Linker}, \& {Arge}}]{Riley15}
{Riley}, P., {Linker}, J.~A., \& {Arge}, C.~N. 2015, Space Weather, 13, 154

\bibitem[{{Riley} {et~al.}(2001){Riley}, {Linker}, \& {Miki{\'c}}}]{Riley01}
{Riley}, P., {Linker}, J.~A., \& {Miki{\'c}}, Z. 2001, \jgr, 106, 15889

\bibitem[{{Riley} {et~al.}(2003){Riley}, {Mikic}, {Linker}, \&
  {Zurbuchen}}]{Riley03}
{Riley}, P., {Mikic}, Z., {Linker}, J., \& {Zurbuchen}, T.~H. 2003, in American
  Institute of Physics Conference Series, Vol. 679, Solar Wind Ten, ed.
  M.~{Velli}, R.~{Bruno}, F.~{Malara}, \& B.~{Bucci}, 79--82

\bibitem[{{Rotter} {et~al.}(2012){Rotter}, {Veronig}, {Temmer}, \& {Vr{\v
  s}nak}}]{Rotter12}
{Rotter}, T., {Veronig}, A.~M., {Temmer}, M., \& {Vr{\v s}nak}, B. 2012,
  \solphys, 281, 793

\bibitem[{{Scherrer} {et~al.}(2012){Scherrer}, {Schou}, {Bush}, {Kosovichev},
  {Bogart}, {Hoeksema}, {Liu}, {Duvall}, {Zhao}, {Title}, {Schrijver},
  {Tarbell}, \& {Tomczyk}}]{Scherrer12}
{Scherrer}, P.~H., {Schou}, J., {Bush}, R.~I., {et~al.} 2012, \solphys, 275,
  207

\bibitem[{{Stone} {et~al.}(1998){Stone}, {Frandsen}, {Mewaldt}, {Christian},
  {Margolies}, {Ormes}, \& {Snow}}]{Stone98}
{Stone}, E.~C., {Frandsen}, A.~M., {Mewaldt}, R.~A., {et~al.} 1998, \ssr, 86, 1

\bibitem[{{Temmer} {et~al.}(2007){Temmer}, {Vr{\v s}nak}, \&
  {Veronig}}]{Temmer07}
{Temmer}, M., {Vr{\v s}nak}, B., \& {Veronig}, A.~M. 2007, \solphys, 241, 371

\bibitem[{{Timothy} {et~al.}(1975){Timothy}, {Krieger}, \&
  {Vaiana}}]{Timothy75}
{Timothy}, A.~F., {Krieger}, A.~S., \& {Vaiana}, G.~S. 1975, \solphys, 42, 135

\bibitem[{{Tu} {et~al.}(2005){Tu}, {Zhou}, {Marsch}, {Xia}, {Zhao}, {Wang}, \&
  {Wilhelm}}]{Tu05}
{Tu}, C.-Y., {Zhou}, C., {Marsch}, E., {et~al.} 2005, Science, 308, 519

\bibitem[{{Verbanac} {et~al.}(2011){Verbanac}, {Vr{\v s}nak}, {Veronig}, \&
  {Temmer}}]{Verbanac11}
{Verbanac}, G., {Vr{\v s}nak}, B., {Veronig}, A., \& {Temmer}, M. 2011, \aap,
  526, A20

\bibitem[{{Vr{\v s}nak} {et~al.}(2007){Vr{\v s}nak}, {Temmer}, \&
  {Veronig}}]{Vrsnak07}
{Vr{\v s}nak}, B., {Temmer}, M., \& {Veronig}, A.~M. 2007, \solphys, 240, 315

\bibitem[{{Wang} \& {Sheeley}(1990)}]{Wang90}
{Wang}, Y.-M., \& {Sheeley}, Jr., N.~R. 1990, \apj, 355, 726

\bibitem[{Wang \& {Sheeley}(1991)}]{Wang91}
Wang, Y.-M., \& {Sheeley}, Jr., N.~R. 1991, \apjl, 372, L45

\bibitem[{{Wang} {et~al.}(1997){Wang}, {Sheeley}, {Phillips}, \&
  {Goldstein}}]{Wang97}
{Wang}, Y.-M., {Sheeley}, Jr., N.~R., {Phillips}, J.~L., \& {Goldstein}, B.~E.
  1997, \apjl, 488, L51

\end{thebibliography}
\end{document}